\documentclass[preprint,aps,prd,groupedaddress,showpacs,floatfix,%
nofootinbib]{revtex4}
\usepackage{epsfig}
\usepackage{hyperref}
\usepackage{graphicx}
\usepackage{dcolumn}
\usepackage{epsf,amsfonts,amssymb}
\usepackage{longtable}
\newcommand{\mathd}{\mathrm{d}}

\newcommand{\mathe}{\mathrm{e}}

\usepackage{relsize}
\usepackage{graphicx,amsmath,amssymb}

\catcode`\à=\active \defà{\`a} \catcode`\À=\active \defÀ{\`A}
\catcode`\á=\active \defá{\'a} \catcode`\Á=\active \defÁ{\'A}
\catcode`\ä=\active \defä{\"a} \catcode`\Ä=\active \defÄ{\"A}
\catcode`\â=\active \defâ{\^a} \catcode`\Â=\active \defÂ{\^A}
\catcode`\å=\active \defå{{\aa}} \catcode`\Å=\active \defÅ{{\AA}}
\catcode`\ç=\active \defç{\c{c}} \catcode`\Ç=\active \defÇ{\c{C}}
\catcode`\è=\active \defè{\`e} \catcode`\È=\active \defÈ{\`E}
\catcode`\é=\active \defé{\'e} \catcode`\É=\active \defÉ{\'E}
\catcode`\ë=\active \defë{\"e} \catcode`\Ë=\active \defË{\"E}
\catcode`\ê=\active \defê{\^e} \catcode`\Ê=\active \defÊ{\^E}
\catcode`\ì=\active \defì{\`{\i}} \catcode`\Ì=\active \defÌ{\`{\I}}
\catcode`\í=\active \defí{\'{\i}} \catcode`\Í=\active \defÍ{\'{\I}}
\catcode`\ï=\active \defï{\"{\i}} \catcode`\Ï=\active \defÏ{\"{\I}}
\catcode`\î=\active \defî{\^{\i}} \catcode`\Î=\active \defÎ{\^{\I}}
\catcode`\ò=\active \defò{\`o} \catcode`\Ò=\active \defÒ{\`O}
\catcode`\ó=\active \defó{\'o} \catcode`\Ó=\active \defÓ{\'O}
\catcode`\ö=\active \defö{\"o} \catcode`\Ö=\active \defÖ{\"O}
\catcode`\ô=\active \defô{\^o} \catcode`\Ô=\active \defÔ{\^O}
\catcode`\ù=\active \defù{\`u} \catcode`\Ù=\active \defÙ{\`U}
\catcode`\ú=\active \defú{\'u} \catcode`\Ú=\active \defÚ{\'U}
\catcode`\ü=\active \defü{\"u} \catcode`\Ü=\active \defÜ{\"U}
\catcode`\û=\active \defû{\^u} \catcode`\Û=\active \defÛ{\^U}
\catcode`\ý=\active \defý{\'y} \catcode`\Ý=\active \defÝ{\'Y}
\catcode`\ÿ=\active \defÿ{\"y} \catcode`\˜=\active \def˜{\"Y}
\catcode`\½=\active \def½{!`} \catcode`\¾=\active \def¾{?`}
\catcode`\ß=\active \defß{{\ss}}

\newcommand{\tmfloatcontents}{}
\newlength{\tmfloatwidth}
\newcommand{\tmfloat}[5]{
  \renewcommand{\tmfloatcontents}{#4}
  \setlength{\tmfloatwidth}{\widthof{\tmfloatcontents}+1in}
  \ifthenelse{\equal{#2}{small}}
    {\ifthenelse{\lengthtest{\tmfloatwidth > \linewidth}}
      {\setlength{\tmfloatwidth}{\linewidth}}{}}
    {\setlength{\tmfloatwidth}{\linewidth}}  \begin{minipage}[#1]{\tmfloatwidth}
    \begin{center}
      \tmfloatcontents
      \captionof{#3}{#5}
    \end{center}
  \end{minipage}}

\usepackage{amsmath,amsfonts,bm}
\usepackage{amssymb}
\usepackage{dcolumn}

\usepackage{graphicx}
\usepackage{epsfig}
\usepackage{hyperref}
\usepackage{graphicx}
\usepackage{dcolumn}

\usepackage{longtable}

\def\babar{\mbox{\slshape B\kern-0.1em{\smaller A}\kern-0.1em
    B\kern-0.1em{\smaller A\kern-0.2em R}}}

\begin{document}
\title{Photon-to-pion transition form factor and pion distribution amplitude from holographic QCD}
\author{Fen Zuo$^{abc}$\footnote{Email: zuof@ihep.ac.cn} and Tao
Huang$^{ab}$\footnote{Email: huangtao@ihep.ac.cn}} \affiliation{
$^a$Institute of High
Energy Physics, Chinese Academy of Sciences, Beijing 100049, China\\
$^b$Theoretical Physics Research Center for Science Facilities,
Chinese Academy of Sciences, Beijing 100049, China\\
$^c$Istituto Nazionale di Fisica Nucleare, Sezione di Bari, Italy}
\begin{abstract}
 We try to understand the recently observed anomalous behavior of the
photon-to-pion transition form factor in the holographic QCD
approach. First the holographic description of the anomalous
$\gamma^*\gamma^*\pi^0$ form factor is reviewed and applied to
various models. It is pointed out that the holographic
identification of the pion mode from the 5D gauge field strength
rather than the gauge potential, as first made by Sakai and
Sugimoto, naturally reproduces the scaling behavior of various pion
form factors. It is also illustrated that in describing the
anomalous form factor, the holographic approach is asymptotically
dual to the perturbative QCD~(pQCD) framework, with the pion mode
$\pi(z)\sim z$ corresponding to the asymptotic pion distribution
amplitude. This indicates some inconsistency in light-front
holography, since $\pi(z)\sim z$ would be dual to $\varphi(x)\sim
\sqrt{x(1-x)}$ there. This apparently contradictory can be attributed to the fact that the
holographic wave functions are effective ones, as observed early by
Radyushkin. After clarifying these subtleties, we employ the relation between the
holographic and the perturbative expressions to study possible asymptotic violation of
the transition form factor. It is found that if one require that the
asymptotic form factor possess a pQCD-like expression, the pion mode
can only be ultraviolet-enhanced by logarithmic factors. The
minimally deformed pion mode will then be of the form $\pi(z)\sim
z\ln (z\Lambda)^{-1}$. We suppose that this deformation may be due
to the coupling of the pion with a nontrivial open string tachyon
field, and then the parameter $\Lambda$ will be related to the quark
condensate. Interestingly, this pion mode leads immediately to
Radyushkin's logarithmic model, which fitted very well the
experimental data in the large-$Q^2$ region. On the other side, the
pQCD interpretation with a flat-like pion distribution amplitude,
proposed by Radyushkin and Polyakov, fails to possess a holographic
expression.
\end{abstract}
\keywords{QCD, AdS-CFT Correspondence}
\pacs{11.25.Tq, 
11.10.Kk, 
11.15.Tk  
12.38.Lg  
} \maketitle

\section{Introduction}
As we know, string theory was born as a candidate of the theory of
strong interaction. The string spectra naturally exhibit the orbital
Regge behavior of hadron trajectories. The large-$N_c$ expansion of
the Feynman diagrams in non-abelian gauge theory shows surprising
similarity to the topological expansion of string scattering
amplitudes~\cite{Hooft1974}. However, phenomenologically the
situation was not so encouraging. The string spectra contain
massless spin-$1$ and spin-$2$ excitations which have no
counterparts in hadron spectra. The hadrons seems to have an
infinite size in the string picture~\cite{Karliner:1988hd}. The tree
level string scattering amplitudes, described by the well-known
Veneziano formula~\cite{Veneziano:1968yb}, exhibit exponential soft
behavior at high momentum, while the observed hadron scattering
amplitudes possess hard power
behavior~\cite{Matveev1973,Brodsky1973}.

A great breakthrough was made in 1997 when Maldacena proposed the
famous anti-de Sitter~(AdS)/CFT correspondence~\cite{Maldacena1998}.
The key point is that the strings describing hadrons live in a
spacetime with some extra warped dimension. This extra dimension
plays the role of energy scale in the gauge theory, and hadrons at
different scales are reflected holographically by strings located at
different positions in the extra direction. With this extra
dimension, many problems in the string description can be readily
solved. For example, the hard power behavior of hadron scattering
amplitudes can be reproduced from the properties of the
corresponding string states in this
dimension~\cite{Polchinski:2001tt}. Furthermore, the underlying
reason for the scaling behavior on the AdS side was found to be very
similar to the Feynman/Drell--Yan
mechanism~\cite{Feynman1972,Drell:1969km}. One can even show that
for specific form factors, the Drell--Yan expression is
asymptotically dual to the corresponding formula in asymptotic AdS
background~\cite{Brodsky:2006uqa}. Based on this, a duality between
the holographic approach and light-cone formalism was established,
with the holographic coordinate $z$ dual to a specific combination
of the longitudinal momentum fraction $x$ and some transverse
distance $b_\perp$~\cite{Brodsky:2006uqa,deTeramond:2008ht}. The
light-cone wave function can then be derived in this so-called
light-front holography framework, from the corresponding field mode
on the AdS side.

However, the original correspondence is valid for exact conformal
field theory at the boundary, and QCD is not a CFT in any case. As a
result, some discrepancy appears when one carefully compares the
power behavior on both sides. On the AdS side the scattering
amplitude scales according to the dimension of the dual boundary
operator~\cite{Polchinski:2001tt}. But in QCD one counts the power
according to the twist of the hadron operator, or the number of hard
constituents inside a hadron~\cite{Lepage:1980fj}. One way out of
this discrepancy is to modify the dictionary of the correspondence,
and replace the conformal dimension of the boundary operator by its
twist~\cite{Brodsky:2006uqa}. This seems to be a little too rough
and artificial, since we have no theoretical support for this on the
AdS side. In QCD the hadrons can never be classified just by the
dimension of the corresponding operators. Instead one should
identify various hadrons according to the corresponding symmetries,
as in~\cite{Son2004,Sakai:2004cn,Erlich:2005qh,DaRold:2005zs}.
Explicit calculations in asymptotic AdS background show that, at
least for the (axial)vector mesons and the pseudoscalar mesons, the
asymptotic behavior of the corresponding form
factors~\cite{Grigoryan:2007my,Grigoryan:2007wn,Grigoryan:2008up}
agrees with the dimensional counting law in QCD and also with
observations~\cite{Matveev1973,Brodsky1973,Radyushkin:2009wx}. The
correct identification of the pion mode in the holographic framework
is crucial to guarantee this agreement. The ultraviolet~(UV)
dominant part of the pion field should come from the 5D gauge
field~\cite{Son2004,Erlich:2005qh}, and should correspond to the
field strength rather than the gauge potential, as first proposed by
Sakai and Sugimoto~\cite{Sakai:2004cn}.

An interesting example is the result for the anomalous
$\gamma^*\gamma^*\pi^0$ form factor, which can be naturally
implemented in the holographic approach by including the
Chern--Simons~(CS) term~\cite{Witten1998a,Son2004,Sakai:2004cn}.
Interestingly, the result at large momentum transfer in asymptotic
AdS background coincides with the leading power perturbative
QCD~(pQCD) expression~\cite{Lepage:1980fj}, if the asymptotic pion
distribution amplitude is
used~\cite{Grigoryan:2008up,Grigoryan:2008cc}. The underlying reason
for this may be attribute to the asymptotic conformal symmetry on
both sides. When one photon is on shell, the corresponding
transition form factor should behave as $1/Q^2$ at large-$Q^2$,
according to the scaling law~\cite{Matveev1973,Brodsky1973}. Thus it
is quite surprising when BABAR's recent data show that the
combination $Q^2 F_{\gamma\pi^0}(Q^2)$ increases continuously with
$Q^2$, rather than approaching a constant~\cite{Aubert:2009mc}. In
the meantime, no such anomalous behavior was observed in the
transition form factors involving $\eta$ and $\eta'$
mesons~\cite{:2011hk}. Thus this anomalous behavior should be
attributed to some specific properties of the pion meson. If one
insists that the experimental data only represent the transition
form factor in the moderate momentum region, then the pQCD approach
with an endpoint-enhanced pion distribution amplitude can describe
the data well~\cite{Chernyak:2009dj,Wu:2010zc,Agaev:2010aq}. On the
other side, it was found in~\cite{Radyushkin:2009zg} that the
increasing pattern can be well approximated by a logarithmic
function of $Q^2$. In the pQCD framework, this logarithmic factor
naturally appear when integrating the hard kernel with a flat-like
pion distribution
amplitude~\cite{Radyushkin:2009zg,Polyakov:2009je}. This kind of
distribution amplitude is completely different from the asymptotic
one, which is much suppressed at the endpoints. Due to this, some
special regularization scheme needs to be introduced in order to
avoid the endpoint singularity~\cite{Radyushkin:2009zg}. Since the
pQCD framework is asymptotically dual to the holographic approach in
describing the $\gamma^*\gamma^*\pi^0$ form factor, one would wonder
if this deformation of the pion distribute amplitude can be made
manifest on the AdS side. Furthermore, one could also try to
understand the observed anomalous behavior independently from the
holographic approach, and to see the corresponding interpretation on
the pQCD side.

The paper is organized in the following way. In the next section we
first review the holographic results of various form factors in a
slice of AdS space, and pay special attention to the anomalous
$\gamma^*\gamma^*\pi^0$ form factor. The asymptotic behavior of the
transition form factor in some other backgrounds were also
calculated and analyzed. In Sec. {I}{I}{I} we show that the minimal
deformation of the holographic pion mode naturally leads to
Radyushkin's logarithmic model, thus describes very well BABAR's
data. The final section is reserved for a short summary.

\section{A brief review of photon-to-pion transition form factor in holographic QCD}
\subsection{Photon-to-pion transition form factor in a hard wall}
First let us review the holographic description of the transition
form factor, which was first done~\cite{Grigoryan:2008up} in the
hard wall model with quark condensate~\cite{Erlich:2005qh}. Later it
was found more natural to work in the non-linear formalism of Son and
Stephanov~\cite{Son2004}, in order to eliminate the infrared~(IR)
boundary term~\cite{Grigoryan:2008up,Grigoryan:2008cc}. So here we
mainly follow the derivation in~\cite{Grigoryan:2008cc}, avoiding to
introduce the infrared counter term.

The {``}open moose" model proposed by Son and Stephanov was
constructed based on non-linear chiral symmetry realization and
large-$N_c$ assumption~~\cite{Son2004}. Under such conditions, one
can introduce infinitely many hidden local symmetries into the
non-linear sigma model. Those fields are finally neatly fitted
together into a single gauge field, propagating in a 5D curved
spacetime. This framework was further developed in
\cite{Sakai:2004cn} and \cite{Hirn:2005nr}. In other words, one can
consider that chiral symmetry breaking is driven by different
boundary conditions in the infrared \cite{Hirn:2005nr}. From the
top-down construction in \cite{Sakai:2004cn}, one can further
attribute this to the coupling in the deep infrared between the
flavor branes and the anti-branes. The left-handed and right-handed
chiral symmetry transformation are identified with the reduced 5D
gauge symmetry at the two ultraviolet boundaries, which should be
considered as the same 4D spacetime but separated from each other in
some extra dimension~\cite{Sakai:2004cn}.

To produce the required anomaly in QCD, the CS term must
be included~\cite{Witten1998a} and we are then dealing with a 5D
Yang-Mills~(YM)+CS theory in a curved spacetime. Following the notation of
\cite{Hirn:2005nr}, we cut the flavor brane from the middle, and
deal with the two sets of gauge fields satisfying different infrared
boundary conditions. To be more specific, we choose to work in the
cut-off AdS spacetime, ie., the hard
wall model~\cite{Polchinski:2001tt,deTeramond:2005su,Erlich:2005qh}. Then
the gauge field satisfies the following equation:
\begin{equation}
\partial_z\left(\frac{1}{z}\partial_zA_\mu(q,z)\right)+\frac{q^2}{z}A_\mu(q,z)=0.\label{eq:vector}
\end{equation}
The UV, $z=0$, normalizable solutions then describe various physical
mesons. Specifically, those vanishing at the IR cut-off $z=z_0$
correspond to the axial mesons, while those with vanishing
derivatives at the cut-off excite vector mesons. For example, the
vector modes are given explicitly by
\begin{equation}
\psi_n^V ( z)   = \frac{\sqrt{2} }{z_0 J_1(\gamma_{0,n})}\, z
J_1(M_{n} z)
\end{equation}
where $ \gamma_{0,n}$ is the $n^{\rm th}$ zero of the Bessel
function $J_0(x)$ and $M_{n} = \gamma_{0,n}/z_0 $. Note that
$\psi_n^V ( z)\sim z^2$ in the UV as expected from the AdS/CFT
dictionary~\cite{Witten1998a,Polyakov1998}. Besides these massive
states, the axial part possesses a massless state, the pion.
Explicitly, one has the axial solution at $q^2=0$
\begin{equation}
\psi^A_0(z)=1-z^2/z_0^2,
\end{equation}
whose derivative $\pi(z)=\partial_z \psi^A_0=-2z/z_0^2$ is UV
normalizable. This is the physical pion in the chiral limit, and
corresponds to the gauge field strength rather than the
potential~\cite{Sakai:2004cn}. Due to this, it has the right UV
behavior, $\pi(z)\sim z$, leading to the observed scaling behavior
for the form
factors~\cite{Grigoryan:2007wn,Grigoryan:2008up,Grigoryan:2008cc}.
Alternatively one can treat the pseudoscalars on the same footing as
the vector/axial modes, then a modification of the conformal
dimension is needed to match the scaling
relation~\cite{Brodsky:2006uqa,Brodsky:2007hb}. From both
approaches, the corresponding pion distribution amplitude has the
form $\varphi_\pi(x)\sim \sqrt{x(1-x)}$ from the light-front
holography relation~\cite{Brodsky:2006uqa,deTeramond:2008ht}.

 Eq.(\ref{eq:vector}) also possesses
non-normalizable solutions, which are dual to the external source in
the boundary theory. In the vector sector, one has
\begin{equation}
\label{JQz} {\cal J} (Q,z) = {Qz}\left[K_1(Qz) + I_1(Qz)
\frac{K_0(Qz_0)}{I_0(Qz_0)} \right],
\end{equation}
where ${\cal J} (Q,z)$ is taken at a spacelike momentum $q$ with
$q^2 =-Q^2$, and satisfies the boundary condition ${\cal J}(Q,0)= 1$
in order to be coupled to the source. It can be shown that ${\cal J}
(Q,z)$ has the following decomposition
formula~\cite{Strassler2004,Grigoryan:2007my}:
\begin{equation}
{\cal J } (Q,z) = g_5\sum_{m = 1}^{\infty}\frac{ F_{m} \psi_m^V(
z)}{ Q^2 + M^2_{m} }.\label{eq:decomposition}
\end{equation}

Various correlation functions can then be derived by differentiating
the YM-CS action with respect to the sources. Further saturating
these correlation functions with the physical states, we obtain the
corresponding form factors. For example, the two-point~(2pt)
correlation function in the vector sector reads
\begin{equation}
\Pi_V(Q^2)=-\frac{1}{g_5^2Q^2} \left[\frac{1}{z}{\cal
J}(Q,z)\partial_z{\cal J}(Q,z)\right]_{z=\epsilon},
\end{equation}
which reproduces exactly the large-$Q^2$ behavior of the quark loop
result~\cite{Shifman1979} if we fix the gauge coupling as
$g_5^2=N_c/12\pi^2$. However, this was shown to be valid only in
holographic models with asymptotic AdS
geometry~\cite{Son2004}.

From the YM action, the $n\to n$ diagonal transition form factors
between the vector mesons were derived to be~\cite{Grigoryan:2007my}
\begin{equation}
F_{nn}(Q^2)=\int^{z_0}_0{\cal J}(Q,z)|\psi^V_n(z)|^2~~\frac{\mathd
z}{z}.
\end{equation}
In the same way, the electromagnetic pion form factor can be derived
and has a similar form~\cite{Grigoryan:2008cc}
\begin{equation}
F_\pi(Q^2)=\frac{1}{g_5^2f_\pi^2}\int^{z_0}_0{\cal
J}(Q,z)|\pi(z)|^2~~\frac{\mathd z}{z},\label{eq:electromagnetic}
\end{equation}
where the different normalization comes from the special kinetic
part of the pion. Since ${\cal J}(Q,z)\sim \mathe ^{-zQ}$ at large
$Q^2$, the asymptotic behavior of the form factors are determined by
the meson modes in the UV region $z\lesssim Q^{-1}$. One easily
finds that $F_{nn}(Q^2)\sim Q^{-4}$ and $F_\pi(Q^2)\sim Q^{-2}$, in
accordance with the perturbative results~\cite{Radyushkin:2009wx}.
As in the 2pt correlation function, the accordance of the power
behavior should also be valid only in asymptotic AdS backgrounds.
This is guaranteed by the same asymptotic symmetry on both sides of
the duality. The scaling behavior in QCD is determined by the
conformal symmetry in the asymptotic free region, which is just the
asymptotic isometry of the dual geometry. So as long as we choose
the proper dual fields, with their effective scaling dimension
coincident with the twists of the hadron states, the same scaling
relations will hold on both sides.

Now let us turn to the anomalous $\gamma^*\gamma^*\pi^0$ form
factor. As is well known, this process is due to the
SU$_A(N_F)$-U$(1)^2$ anomaly, see e.g ~\cite{Son2004}, which is
manifested in the dual theory through the Chern-Simons interaction .
From the corresponding term one can derive this form factor in the
same way as the previous ones, and the result
is~\cite{Grigoryan:2008cc}
\begin{equation}
F_{\gamma^*\gamma^*\pi^0}(Q_1^2,Q_2^2)=-\frac{N_c}{12\pi^2f_\pi}\int^{z_0}_0{\cal
J}(Q_1,z){\cal J}(Q_2,z)\pi(z) \mathd z,\label{eq:anomalous}
\end{equation}
where the normalization is fixed by the QCD axial anomaly at
$Q_1^2=Q_2^2=0$. In the special case with one photon onshell, we get
the transition form factor
\begin{equation}
F_{\gamma\pi^0}(Q^2)=-\frac{N_c}{12\pi^2f_\pi}\int^{z_0}_0{\cal
J}(Q,z)\pi(z) \mathd z,\label{eq:transition}
\end{equation}
Notice that due to the topological character of the CS action, no
metric factor appears explicitly in the above form factors.
Moreover, from the equation of motion for $\psi_0^A(z)$ one can
easily show that the transition form factor coincides with the
electromagnetic pion form factor up to the normalization factor
$\frac{N_c}{12\pi^2f_\pi}$~\cite{Grigoryan:2008cc,Stoffers:2011xe}.
Denoting $Q_1^2=(1+\omega)Q^2$ and $Q_2^2=(1-\omega)Q^2$, the
normalized form factor
 $K(Q_1^2,Q_2^2)=\frac{12\pi^2f_\pi}{N_c}
 F_{\gamma^*\gamma^*\pi^0}(Q_1^2,Q_2^2)$ has the following
 asymptotic behavior
 \begin{equation}
K(Q_1^2,Q_2^2)\to \frac{s_0}{Q^2}\int^1_0 \frac{6x(1-x)\mathd
x}{1+\omega(2x-1)}, \label{eq-pQCD}
\end{equation}
where $s_0=8\pi^2f_\pi^2$. Surprisingly, this coincides with the
pQCD result with the asymptotic pion distribution amplitude
$\varphi_\pi^{\rm{as}}(x)=6x(1-x)$~~\cite{Lepage:1980fj}. For the
convenience of later discussions, let us repeat some details of the
derivation of the above expression~\cite{Grigoryan:2008up}. First
note that the Bessel function $K_1$ can be represented through the
expression
\begin{equation}
 \chi K_1(\chi)=\int^\infty_0 \mathe^{-\chi^2/4u-u} \mathd u.
\end{equation}
Substituting this in eq. (\ref{eq:anomalous}) and integrating over
$\chi$ one finds
\begin{equation}
K(Q_1^2,Q_2^2)\to \frac{s_0}{Q^2}\int_0^\infty \int_0^\infty
\frac{u_1u_2\mathe^{-u_1-u_2}\mathd u_1\mathd
u_2}{u_2(1+\omega)+u_1(1-\omega)}.
\end{equation}
Further defining $u_2=x\lambda$ and $u_1=(1-x)\lambda$ and
integrating over $\lambda$, one finally obtain Eq. (\ref{eq-pQCD}).
A crucial property that guarantees this asymptotic form is the
UV scaling of the pion mode, $\pi(z)\sim z$. If the pion function
scales as $z^n$, then the whole integrand in Eq. (\ref{eq-pQCD}),
together with the $Q^{-2}$ factor, will develop an overall power
$(n+1)/2$. So only when $n=1$ the pQCD result can be retained.
Correspondingly, the asymptotic behavior for the photon-pion
transition form factor is the same as the pQCD result with the
asymptotic pion distribution amplitude. However, the mechanism from
both sides are rather different. The scaling behavior in the pQCD
part is completely determined by the hard kernel, no matter what the
distribution amplitude is. In the holographic side, the power
behavior appears only after we integrate the meson mode folded with
the source profile, and strongly depends on the UV behavior of the
meson solution. This is in much common with the soft contributions
in the Light-Cone Sum Rule~(LCSR) approach, as shown
in~\cite{Braun:1994ij,Khodjamirian:1997tk}.

Now some discrepancy between light-front holography and the
above result appears. From the previous discussion on the
electromagnetic pion form factor, the pion mode $\pi(z)\sim z$ would
correspond to a distribution amplitude $\varphi(x)\sim
\sqrt{x(1-x)}$ from the light-front holography. But here we find
that in the large-$Q^2$ limit the holographic expression with
$\pi(z)\sim z$ is exactly dual to the pQCD result with the asmptotic
pion distribution amplitude in the asymptotic form. This kind of
discrepancy can also be found from the study of the $\rho\to \pi$
transition form factor from both the holographic approach and the
light-cone sum rules~\cite{Zuo:2009hz}. To ensure the required
$Q^{-4}$ scaling in the UV, we again should properly define the pion
mode and find $\pi(z)\sim z$. On the other side, it has been long
known in the light-cone sum rules framework that the soft
contributions give rise to a $Q^{-4}$ contribution only when the
pion distribution amplitude vanishes linearly at the endpoints, the
same as the asymptotic one~\cite{Braun:1994ij}. Based on this, it
was speculated that light-front holography would generally fail
when the result depends linearly on the distribution
amplitude~\cite{Radyushkin:2009wx}.

To see how this contradiction comes out we must check the derivation
in light-front holography~\cite{Brodsky:2006uqa} carefully. There
was a crucial step toward the final holographic expression for the
light-cone wave function. Namely, one must replace the original
single particle density by an effective two-body one. But by
definition, the single particle density receives contributions from
the whole tower of Fock states. Therefore, the light-cone wave
functions so obtained are only effective ones, and contain
contributions from all light-cone states. This fact has already been
pointed out in ref.~\cite{Radyushkin:2006iz} since light-front
holography was advocated. It was shown that the holographic wave
function normalizes to unity rather than the valence Fock state
probability, and exhibits unusual power behavior for large
transverse momentum. Therefore, one must be very careful when
applying the holographic distribution amplitude in other approaches,
such as the perturbative calculation and light-cone sum rules.
On the other hand, one would wonder if we can separate the
contributions from different light-cone Fock states. The above
result for the $\gamma^*\gamma^*\pi^0$ form factor gives us the
answer. Generally, different Fock states contribute with different
powers of the large momentum involved. From the leading power
contribution one would find the valence-state wave function, exactly
what one find in the $\gamma^*\gamma^*\pi^0$ and
$\rho^0\gamma^*\pi^0$ form factors discussed above. The other power
contributions in the large-$Q^2$ expansion should give us the wave
functions of other Fock states, respectively. A drawback of this
separation is that one can only obtain wave functions in the
asymptotic region, as recently commented in~\cite{Brodsky:2011xx}.
It would be interesting to see if this procedure could be continued
to moderate and even lower momentum region.


\subsection{Photon-to-pion transition form factor in other holographic models}

In the previous section we have reviewed the derivation of the
$\gamma^*\gamma^*\pi^0$ form factor in the hard wall model, and
shown that the asymptotic form factor coincides with that in pQCD.
In this section we will derive the asymptotic transition form factor
in other holographic models, to see how the results change when the
backgrounds vary. In particular, we want to confirm that only in
asymptotic AdS backgrounds the scaling behavior on the holographic
side coincides with the perturbative prediction.

First let us consider the {``}cosh" background proposed
in~\cite{Son2004}, which corresponds to a smooth connection of two
IR-cut-off AdS spacetime slices. This background can be conveniently
expressed using the coordinate $u\sim -\log z$ as
\begin{equation}
\mathd s^2=-\mathd u^2+\Lambda^2 \cosh ^2u\,\eta_{\mu\nu}\mathd
x^\mu\mathd x^\nu,
\end{equation}
from which follows the equation of motion
\begin{equation}
\partial_u\left[\cosh^2u\,\partial_uA_\mu(q,u)\right]+\frac{q^2}{\Lambda^2}A_\mu(q,u)=0.
\end{equation}
The explicit solution for the propagator ${\cal J}(Q,u)$ was derived
in~\cite{Son:2010vc}. Here we just need to notice that ${\cal
J}(Q,u)$ depends only on the combination $Q\mathe^{-u}$ in the UV
region. The pion wave function is easily found to be
$\pi(u)=\cosh^{-2}u$. So the asymptotic transition form factor reads
\begin{eqnarray}
F_{\gamma\pi^0}(Q^2)&\sim&\int^{\infty}_0{\cal J}(Q,u)\,\pi(u)\,
\mathd
u,\nonumber\\
&\sim&\int^{\infty}_0{\cal J}(Q\mathe^{-u})\,\mathe^{-2u}\, \mathd
u,\nonumber\\
&\sim &Q^{-2} \int^{\infty}_0{\cal J}(zQ)\, \mathd (z^2Q^2)\,,
\end{eqnarray}
with the expected power behavior.

Another simple example is the flat spacetime. Actually the result in
this background has been obtained in~\cite{Cappiello:2010uy}, but in
a different framework~\footnote{Note that there should be a minus
instead of plus sign in the expression for the bulk-to-boundary
propagator derived in~\cite{Cappiello:2010uy}. Due to this mistake,
the result for the transition form factor there is different from
ours.}. In this case we have to introduce a UV cut-off to obtain finite
results~\cite{Son2004}. One can easily finds that
\begin{equation}
{\cal J}(Q,r)=\frac{\cosh rQ}{\cosh r_0Q},
\end{equation}
and $\pi(r)= 1/r_0$. So the transition form factor reads
\begin{eqnarray}
F_{\gamma\pi^0}(Q^2)&=& r_0^{-1}\,\int^{r_0}_0{\cal J}(Q,r)\, \mathd
r\nonumber\\
&=& (r_0Q\cosh r_0Q)^{-1}\,\int^{r_0Q}_0 \cosh rQ \, \mathd
(rQ)\nonumber\\
&=& \frac{\tanh r_0Q}{r_0Q},
\end{eqnarray}
which vanishes slower than $Q^{-2}$.

Finally let us consider Sakai and Sugimoto's construction based on
intersected $D4$-$D8$/$\bar{D}8$ branes. The transition form factor
in this model has been studied recently
in~\cite{Cappiello:2010uy,Stoffers:2011xe}. The flavor part action
reads
\begin{equation}
S=\kappa\int \mathd^4x\mathd r {\rm
tr}\left[-\frac{1}{2}K^{-1/3}\,F^2_{\mu\nu}+K\,F^2_{\mu r}\right],
\end{equation}
where $\kappa=\frac{\lambda N_c}{216\pi^3}$ with $\lambda$ the 't
Hooft coupling, and $K(r)=1+r^2$. The equation of motion follows
directly
\begin{equation}
K^{1/3} \partial_r (K\,\partial_r
V(q,r))+\frac{q^2}{M_{\rm{KK}}^2}V(q,r)=0,
\end{equation}
with $M_{\rm{KK}}$ the only energy scale in the model. In the UV
region $r\to \pm \infty$, one has the asymptotic solution ${\cal
J}(Q,r)\sim {\cal J}(r/Q^3)$ and $\pi(r)\sim (1+r^2)^{-1}$. The
transition form factor then has the form
\begin{eqnarray}
F_{\gamma\pi^0}(Q^2)&\sim& \int^\infty_0 {\cal
J}(r/Q^3)\,(1+r^2)^{-1}\,\mathd r\nonumber\\
&\sim& Q^{-3}\,\int^\infty_0 \,{\cal J}(r/Q^3)\,
(r/Q^3)^{-2}\,\mathd (r/Q^3),
\end{eqnarray}
which vanishes faster than those in asymptotic AdS backgrounds. This
has been explicitly shown using the decomposition formula
in~\cite{Stoffers:2011xe}.

\section{Holographic implementation of BABAR's observation and the logarithmic model}
\subsection{Minimal deformation of the pion mode}
From the above discussion we have seen that for asymptotic AdS
backgrounds, the holographic description predicted the $Q^{-2}$
behavior for the transition form factor, the same as the
perturbative result.  This was also supported by the CLEO results
roughly in the interval
$2\,\rm{GeV}^2<Q^2<10\,\rm{GeV}^2$~\cite{Gronberg:1997fj}. It was
quite unexpected when the BABAR Collaboration reported their recent
data up to $Q^2\sim 40\, \rm{GeV}^2$, which shows that the form
factor decreases much slower than the predicted $Q^{-2}$ power.
Therefore, how should we modify the holographic description, in
order to produce the observed behavior of the transition form
factor?


From the discussion in last section we already know that the
asymptotic form factor is determined by the explicit form of the
propagator and the UV scaling of the pion mode, which are in turn
determined by the background metric. To keep all the other
correlation functions and form factors unchanged, we will insist on
that almost all the fields still live in an asymptotic AdS
spacetime. Only the pion field feels some special kind of effective
metric. This may be due to its coupling to the open string tachyon
field, which is believed to be responsible for dynamical chiral
symmetry
breaking~\cite{Erlich:2005qh,DaRold:2005zs,Casero:2007ae,Bergman:2007pm,Dhar:2007bz,Aharony:2008an}.
One may wonder why the other (pseudo)goldstone bosons, e.g., the
$\eta$ and $\eta'$ mesons, don't feel this kind of specific
background metric. Since the U$_{\rm{A}}(1)$ symmetry is explicitly
broken by the anomaly, $\eta'$ is rather different from the $\pi$
meson, and thus may be less affected by the tachyon. The $\eta$
meson indeed couples with the tachyon, but mainly to the quark mass
part due to the large strange quark mass, while for $\pi$ the quark
condensate part of the tachyon profile dominates. One may use these
arguments to explain why no violation of the scaling behavior in the
transition form factors of the $\eta$ and $\eta'$ is
observed~\cite{:2011hk}.

Now we will try to reproduce the observed enhancement of the
transition form factor from the holographic expression
(\ref{eq:transition}). We maintain the form of the propagator ${\cal
J}(Q,z)$ and only modify the UV behavior of the pion solution. As
emphasized in the previous section, the scaling $\pi(z)\sim z$ is
necessary to guarantee a dual pQCD expression. However, logarithmic
factors are still allowed. In order to get an enhanced form factor
from eq. (\ref{eq:transition}), the pion mode must be enhanced in
the UV region. Thus the minimal deformation of the pion function
takes the form
\begin{equation}
\pi(z)=C s_0\, z \ln (\Lambda z)^{-1},\,\,, 0<z<\epsilon \ll
z_0.\label{eq:pion}
\end{equation}
Here $\Lambda$ is of energy
dimension, and the constant $C$ will be dimensionless since we have included a factor of $s_0$. According to our naive guess, $\Lambda$ would be related to the quark condensate, and thus will be roughly of a few hundreds
MeV. Then the transition from factor becomes asymptotically
\begin{eqnarray}
Q^2F_{\gamma\pi^0}(Q^2)&\approx& -2Cf_\pi\, Q^2\,\int^\epsilon_0\,z \ln
(\Lambda z)^{-1} \,
zQK_1(zQ)\, \mathd z\nonumber\\
&\approx&-2Cf_\pi\,\int^\infty_0 \chi^2\,(\ln
\frac{Q}{\Lambda}-\ln \chi)\,K_1(\chi)\, \mathd \chi \nonumber\\
&=&-2C\,f_\pi\, \ln \frac{Q^2}{3.43
\Lambda^2}.\label{eq:transition2}
\end{eqnarray}

Having fixed the asymptotic form of the pion field, the effective
metric function in the UV region can in turn be determined. To do
this let us express the effective metric in the following form
\begin{equation}
\mathd s^2=\frac{1}{h^2(z)z^2}(-\mathd z^2+\eta_{\mu\nu}\mathd
x^\mu\mathd x^\nu).
\end{equation}
Then the equation for the zero mode of the axial field becomes
\begin{equation}
\partial_z\left(\frac{1}{h(z)z}\partial_z\psi^A_0(z)\right)=0,
\end{equation}
together with the boundary condition $\psi^A_0(0)=1$ and
$\psi^A_0(z_0)=0$. Moreover, $\psi^A_0(z)$ should satisfy the
normalization condition
\begin{equation}
\frac{1}{g_5^2}\int^{z_0}_0\,\frac{\mathd z}{z h(z)}(\partial_z
\psi^A_0(z))^2=f_\pi^2,
\end{equation}
in order to ensure the correct kinetic action of the pion. Using the
equation of motion this translates into the boundary condition
\begin{equation}
\frac{1}{g_5^2}\left(\frac{1}{h(z)z}\partial_z
\psi^A_0(z)\right)|_{z=0}=f_\pi^2.
\end{equation}
Since the pion field is directly related to $\psi^A_0$ as
$\pi(z)=\partial_z \psi^A_0(z)$, one immediately read off the
asymptotic form of $h(z)$
\begin{equation}
h(z)=\frac{2}{3} \ln (\Lambda z)^{-1}, \,\, z\to 0.
\end{equation}
It will be interesting to see how the tachyon field can provide such
a modification factor to the effective metric of the pion. Moreover,
in the above derivation we have only assumed that the quadratic
action of the pion involves the deformed metric. We can further
assume that the cubic terms involving two pions feel the same
metric. Then one finds that the electromagnetic pion form factor
$F_\pi(Q^2)$ is again identical to the normalized transition form
factor $K(0,Q^2)$, and thus has the same asymptotic behavior. One
can directly check this by taking the large-$Q^2$ limit of
Eq.~(\ref{eq:electromagnetic}), with $1/z$ replaced by $1/zh(z)$,
and $\pi(z)$ given by Eq.~(\ref{eq:pion}).

As a byproduct, we can further study the $\rho^0\to \pi^0$
transition form factor, which can be derived from
eq.~(\ref{eq:anomalous}) by using the decomposition relation
(\ref{eq:decomposition})~\cite{Zuo:2009hz}:
\begin{equation}
F^{\rho^0\pi^0}(Q^2)= -\frac{N_c}{12\pi^2 f_\pi}
\frac{g_5m_\rho}{2}\int_0^{z_0}{\cal J}(Q,z) \,\psi_1^V (
z)\,\pi(z)\mathd z . \label{eq:ff}
\end{equation}
Substituting the pion mode (\ref{eq:pion}), one finds that at
large-$Q^2$ region it acquires an analogous logarithmic enhancement
\begin{equation}
F^{\rho^0\pi^0}(Q^2)\to \frac{8\sqrt{2}\pi^2}{3}f_\pi f_\rho
m_\rho^2 Q^{-4} \ln \frac{Q^2}{15.4 \Lambda^2}.
\end{equation}
As mentioned previously, the holographic description of this
asymptotic form factor is very similar to the traditional LCSRs. So
it will be interesting to see if such an anomalous behavior can be
obtained in the LCSRs. This can be done in a parallel way as above.
In this case we have to deform the pion distribution amplitude at
the endpoints as
\begin{equation}
\varphi(x)\sim x(1-x)\ln [Dx(1-x)]^{-1},\label{eq:pionDA2}
\end{equation}
with $D$ some dimensionless constant. Then the asymptotic form
factor from the corresponding
LCSR~\cite{Khodjamirian:1997tk,Zuo:2009hz} reads
\begin{equation}
F^{\rho^0\pi^0}_{\rm{LC}}(Q^2)\sim \frac{f_\pi}{f_\rho}
\exp\left[\frac{m_\rho^2}{M_{\rm{B}}^2}\right]
\frac{M_{\rm{B}}^4}{Q^4} \left[\ln
\frac{Q^2}{M_{\rm{B}}^2}+\rm{constant}\right],
\end{equation}
where $M_{\rm{B}}^2$ is the Borel parameter. The interval of $M_B^2$
was shown to shifting from $0.9-1.6~ \rm{GeV}^2$ at $Q^2\sim 1
~\rm{GeV}^2$ to $0.5-0.9 ~\rm{GeV}^2$ at $Q^2=10~
\rm{GeV}^2$~\cite{Khodjamirian:1997tk}. Thus, taking the
undetermined constant into account, $M_{\rm{B}}^2$ may be roughly of
the same order of magnitude as $\Lambda^2$.

\subsection{Experimental fit and the logarithmic model}
\begin{figure}[htbp]
\centerline{
\epsfig{file=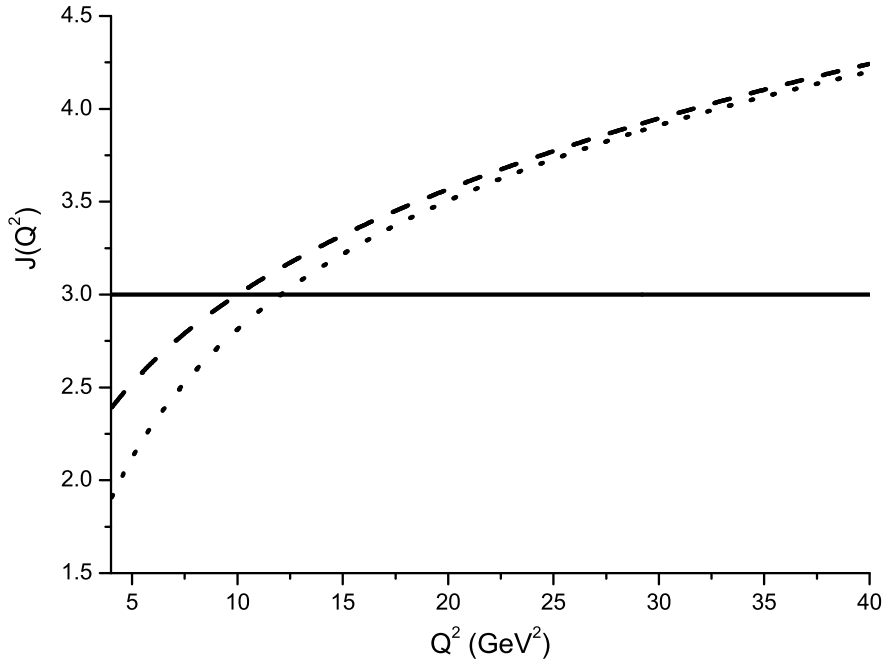,width=0.9\textwidth
,clip=} } \caption{\it Our result for the function $J(Q^2)$~(dotted line) in comparison with the experimental fit (\ref{eq:experiment}) (dashed line). The asymptotic perturbative prediction $J_{\rm{as}}=3$ is also plotted (solid line).}\label{fig:TFF}
\end{figure}

The observed data for the form factor in the range
$4\,\rm{GeV}^2<Q^2<40\,\rm{GeV}^2$ can be well fitted
as~\cite{Aubert:2009mc}
\begin{equation}
Q^2F_{\gamma\pi^0}(Q^2)\cong 2f_\pi
\left(\frac{Q^2}{10\,\rm{GeV}^2}\right)^{0.25}.\label{eq:experiment}
\end{equation}
Meanwhile, the experimental data can also be well fitted by the
logarithmic model~\cite{Radyushkin:2009zg}\footnote{See also
\cite{Dorokhov:2009dg} for a double logarithmic fit of the data,
which comes from the quark triangle loop.}
\begin{equation}
Q^2F_{\gamma\pi^0}(Q^2)\cong
\frac{2f_\pi}{3}\,\ln\left(\frac{Q^2}{M^2}+1\right),\label{eq:logrithmic}
\end{equation}
which almost coincides with the previous fit in the interval
$15\,\rm{GeV}^2\lesssim Q^2<40\,\rm{GeV}^2$ if we take
$M^2=0.6\,\rm{GeV}^2$. Asymptotically, this is just eq.
(\ref{eq:transition2}) from our minimal deformation in the
holographic approach. Equating the two expressions asymptotically,
one finds
\begin{equation}
C=-1/3,\,\Lambda^2=M^2/3.43.
\end{equation}
For $M^2=0.6~\rm{GeV}^2$, this gives $\Lambda=0.42~\rm{GeV}$. This
is roughly the order of the cube root of the quark condensate, which can be introduced through
the vacuum solution of the tachyon field as in refs.~\cite{Erlich:2005qh,DaRold:2005zs} with
the $N_c$ factor taken into account~\cite{Cherman:2008eh,Jugeau:2009mn}. We plot our result (\ref{eq:transition2}) with these
parameters in Fig. \ref{fig:TFF}, where the function $J(Q^2)$ is defined $$J(Q^2)=3Q^2F_{\gamma\pi^0}(Q^2)/2f_\pi.$$ Just as in ref.~\cite{Radyushkin:2009zg}, our result is very close to the experimental fit (\ref{eq:experiment}) in the region $15\,\rm{GeV}^2\lesssim Q^2<40\,\rm{GeV}^2$. When $Q^2$ is smaller than about $15\,\rm{Gev}^2$, our result starts to deviate much from the experiment fit. This is as expected, since we only specify the effective metric for the pion in the near UV region. Certainly one can extend this effective metric to the full
spacetime and obtain the form factor in the whole momentum region. However, this is somehow beyond the scope of the present study, and will be left
for future work.

How to understand these results? Should we take them as the true
asymptotic behavior of the transition form factor? Or they just
reflect some kind of anomalous behavior at moderate $Q^2$, and
finally the form factor will return to their $Q^{-2}$ behavior later
at higher momentum transfer? If it is the latter, we just need to
modify some details of various framework to fit the data. For
example, in the pQCD approach the endpoint-enhanced CZ~(Chernyak--Zhitnitsky)-like distribution amplitude
of pion can describe the
data much better than the asymptotic
one~\cite{Chernyak:2009dj,Wu:2010zc,Agaev:2010aq}. On the
holographic side, truncating the propagator to the lowest $\rho$
resonance also seems to push the prediction close to experimental
values~\cite{Stoffers:2011xe}. However, in this paper we will assume
that the BABAR results reflect the true asymptotic behavior of the
transition form factor, and this behavior is captured by
Eq.(\ref{eq:logrithmic}). Then it seems very difficult to understand
this anomalous behavior from the traditional approaches, either
perturbative calculations ~\cite{Brodsky:2011yv} or the LCSR
framework~\cite{Bakulev:2011rp}.

Interestingly, this kind of logarithmic-enhanced behavior can be
naturally produced with the flat-like pion distribution amplitude in
the perturbative approach~\cite{Radyushkin:2009zg,Polyakov:2009je}.
However, some modification needs to be done in the framework, since
the flat function cannot compensate the endpoint singularity in the
hard kernel any longer. A natural way to eliminate the singularity
is to retain the transverse momentum part in the quark propagator of
the hard kernel~\cite{Radyushkin:2009zg,Li:2009pr}, which has been
employed intensively in the literature. Alternatively, one can
retain the transverse momentum dependence in the pion wave function,
which after integration up to the interaction scale leads to an
effective suppression of the endpoint
contributions~\cite{Radyushkin:2009zg}.

\subsection{Relation to perturbative formalism}
Now let us see what is the dual perturbative expression for our
holographic result, extending the previous duality relation in
asymptotic AdS backgrounds~\cite{Grigoryan:2008up}. To do this, one
substitutes our deformed pion mode (\ref{eq:pion}) into the general
expression for the $\gamma^*\gamma^*\pi^0$ form factor
(\ref{eq:anomalous}). Following the procedure
in~\cite{Grigoryan:2008up}, one can transform it into a pQCD-like
expression:
\begin{eqnarray}
K(Q_1^2,Q_2^2)\to\frac{s_0}{3}\int^1_0&&\frac{6x(1-x)}{xQ_1^2+(1-x)Q_2^2}\nonumber\\
&&\left[\frac{1}{3}\ln\frac{xQ_1^2+(1-x)Q_2^2}{5.65x(1-x)\Lambda^2}\right]\mathd
x.\label{eq:pQCD2}
\end{eqnarray}
As a direct check of this, one can take one photon on shell and
reproduce exactly Eq. (\ref{eq:transition2}). When both photons take
the same momentum square, the form factor can be found to behave as
\begin{equation}
Q^2F_{\gamma^*\gamma^*\pi^0}(Q^2,Q^2)\to \frac{2f_\pi}{9}\ln
\frac{Q^2}{1.07\Lambda^2}.
\end{equation}
Interestingly, the same behavior was also found in the constituent
chiral quark model~\cite{deRafael:2011ga}, with the constituent
quark mass playing the role of $\Lambda$.

Comparing to the derivation in asymptotic AdS
backgrounds~\cite{Grigoryan:2008up}, the only difference comes from
the deformation of the pion wave function, which was transformed
into the additional logarithmic factor shown in the square bracket
of Eq.~(\ref{eq:pQCD2}). One may further attribute this factor to
the transverse momentum part of the light-cone pion wave function,
in the same spirit as the soft exponential factor
in~\cite{Radyushkin:2009zg}. However, this interpretation is
slightly acceptable, since this factor does not converge in the
large-$Q^2$ limit to the expected $\ln x(1-x)$ factor in the
distribution amplitude (\ref{eq:pionDA2}), as deduced from the LCSR
analysis. Thus one may suspect that there are some non-factorizable
contributions, which after integrating over the transverse momentum
give the logarithmic factor.

One may reverse the duality to see if the pQCD interpretation with a
flat pion DA~\cite{Radyushkin:2009zg,Polyakov:2009je} has a
holographic description. Unfortunately, the answer is no. This is
because in the holographic derivation the asymptotic DA and the hard
kernel are always bound together, as long as the source mode ${\cal}
J(Q,z)$ propagates in asymptotic AdS background. This should be so,
or all the other form factors will deviate from the observed scaling
behavior. On the other hand, the experimental fit
(\ref{eq:experiment}) does have a holographic description. In this
case one has to deform the pion mode to be proportional to
$\sqrt{z}$. However, then one finds that the perturbative
interpretation is lost: the hard kernel has a fractional overall
power instead. In fact, one can rigorously prove that, the pion
distribution amplitude in the pQCD expression must be of the
asymptotic form, in order to have a holographic dual description.
Correspondingly, the holographic pion mode at small $z$ must be
linear of $z$, with only logarithmic factors permitted, in order to
have a pQCD interpretation.

\section{Summary}
In this paper we attempt to obtain a holographic interpretation of
the anomalous asymptotic behavior of the photon-to-pion transition
form factor, recently observed at BABAR. It is found that this can
be naturally implemented if the pion field develops an additional
logarithmic factor in the ultraviolet region. After performing the
integration involving the source mode, this factor turns into the
logarithmic enhancement of the form factor, which was shown to
describe the BABAR data quite well in the large-$Q^2$ region. When
this description is converted into a pQCD expression asymptotically,
one finds a direct logarithmic factor in addition to the hard kernel
and the asymptotic pion DA. In the perturbative picture this factor
may come from some non-factorizable contributions, which could be
due to the transverse momentum entanglement of the hard kernel and
the pion. On the other side, the pQCD interpretation with a
flat-like pion DA does not have a counterpart in the holographic
framework. We also guess that this deformation of the pion mode may
result from the coupling with the background tachyon field. This is
natural, since the open string tachyon field was considered to
responsible for both explicit and dynamical chiral symmetry
breaking. Therefore, if the observed violation is confirmed by
future experiments, it can help us to constrain the potential
function of the tachyon field.

\hspace{1cm}

{\bf Acknowledgements}: We are grateful to Stanley Brodsky for very
useful communications. This work was supported in part by Natural
Science Foundation of China under Grant No.~10975144, No.~10735080.

\newpage

\end{document}